\documentclass[%
 reprint,
 amsmath,amssymb,
 aps,
floatfix,
]{revtex4-2}

\usepackage{graphicx}
\usepackage{dcolumn}
\usepackage{bm}  
\usepackage[dvipsnames]{xcolor}
\usepackage{hyperref}
\usepackage{natbib}
\usepackage{comment}
\usepackage{subcaption}
\usepackage{ragged2e}

\hypersetup{
    unicode=false,          
    pdftoolbar=true,        
    pdfmenubar=true,        
    pdffitwindow=false,     
    pdfstartview={FitH},    
    pdfauthor={A Dwivedi et al},     
    colorlinks=true,       
    linkcolor=NavyBlue,          
    citecolor=PineGreen,        
    urlcolor=OrangeRed,     
    pdfborder={0 0 0},
}

\begin{document}
	
	\title{Emergence of Active Foams in a Minimal Cognitive Flocking Model}

\author{Aryan Dwivedi\textsuperscript{$\star$}}  
\email{ms21204@iisermohali.ac.in}
\affiliation{Department of Physical Sciences,
	Indian Institute of Science Education and Research Mohali,
	Sector 81, Knowledge City, Manauli 140306, India}    

\author{Diganta Bhaskar\textsuperscript{$\star$}} 
\email{ph23060@iisermohali.ac.in}
\affiliation{Department of Physical Sciences,
	Indian Institute of Science Education and Research Mohali,
	Sector 81, Knowledge City, Manauli 140306, India}    

\author{Anil Kumar Dasanna}
\thanks{Corresponding author}
\email{adasanna@iisermohali.ac.in}
\thanks{\\$\star$ These authors contributed equally.}
\affiliation{Department of Physical Sciences,
	Indian Institute of Science Education and Research Mohali,
	Sector 81, Knowledge City, Manauli 140306, India}

\begin{abstract}
We investigate a minimal visual perception-driven model of self-propelled particles that interact through purely position-based cohesive steering within a finite vision cone, in the absence of velocity alignment and physical volume exclusion. Building on a previously proposed minimal cognitive model, we extend the interaction range from the previously studied short-range and distance-independent limits to an intermediate regime, enabling a systematic investigation of how a finite vision range influences collective behavior. We find that this regime exhibits a rich state diagram that combines characteristics of both limiting cases while giving rise to qualitatively new collective states. Besides the gas, nematic-band, and polar file states, we uncover the emergence of \emph{active foams} -- system-spanning fractal-like networks sustained by dynamically evolving particle streams. These structures arise solely from non-reciprocal vision-based interactions and are absent in both the short and long vision-range limits. For larger vision angles, the system self-organizes into a variety of aggregate morphologies, including condensates, rings, and multi-twist filaments, revealing a strong interplay between interaction range and anisotropic perception in shaping collective organization. 
\end{abstract}

\maketitle

\section{Introduction}

Self-propelled particles have attracted considerable attention over the past few decades as minimal models for investigating nonequilibrium phenomena at both individual and collective levels in synthetic as well as living systems. The principal distinguishing feature of active matter from conventional driven out-of-equilibrium systems is its ability to break detailed balance--a hallmark of equilibrium systems--continuously at the level of individual constituents, thereby violating time-reversal symmetry. Active particles continuously convert energy from internal or environmental sources (e.g., ATP hydrolysis in living organisms) into persistent mechanical motion, maintaining the system intrinsically out of equilibrium. In most of the \emph{dry} active systems studied, momentum conservation is additionally broken, as the particles move through a resistive background medium, which also justifies the constancy of individual velocities and manifests in the breaking of the Galilean invariance~\cite{te2026colloquium, ramaswamy2010, de2015introduction, tailleur2022active}.

On the theoretical side, active Brownian particles--isotropic particles undergoing persistent self-propulsion together with stochastic rotational diffusion--provide simple yet strong frameworks for studying emergent collective phenomena in active matter such as flocking, motility-induced phase separation (MIPS), and related phenomena~\cite{vicsek2012, marchetti2013, bechinger_active_2016}. Some of these coordinated behaviors often emerges through a rapid transition from a disordered state to a highly aligned moving phase, as first demonstrated in the Vicsek model~\cite{vicsek1995}. Similar collective phenomena occur across a broad range of biological systems, from bacterial colonies and biofilms to schools of fish, flocks of birds, herds of mammals and human crowds~\cite{buhl2006disorder, bonabeau1997self, Camazine2003SelfOrganization, xavier2019single, shaebaniComputationalModelsActive2020, romanczukActiveBrownianParticles2012, elgetiPhysicsMicroswimmersSingle2014, helbing2000simulating, warren2018collective}. Similar spectacle are also observed in synthetic and engineered systems, including active colloids, vehicular traffic, drone swarms~\cite{illien2017fuelled, PhysRevE.89.062316, Cavagna_review2014, verdoucq2025flocking}. Remarkably, these large-scale coordinated and self-organized patterns emerge solely from local interactions among neighboring agents, without requiring any centralized control.

The Vicsek model demonstrated that local velocity alignment between neighboring particles is sufficient to produce coherent flocking, establishing alignment as a fundamental mechanism for collective motion~\cite{vicsek1995, PhysRevLett.92.025702}. However, a wide variety of collective states, including flocking, milling, and coherent aggregation, can emerge even in the absence of explicit velocity alignment~\cite{romanczuk2009, strombom2011, moussaid2011, pearce2014, huepe2013, huepe2015, barberis2016, grossmann2013, soto2014}. In vision-based self-propelled systems, with perception restricted to a finite field of view, purely attractive or cohesive interactions are sufficient to generate such emergent dynamics. This limited perception renders the interactions intrinsically nonreciprocal, as one particle may perceive another without being perceived in return, thereby breaking action--reaction symmetry~\cite{barberis2016}. More generally, nonreciprocal interactions have recently emerged as a unifying paradigm in active matter, giving rise to a broad spectrum of nonequilibrium phenomena, including pattern formation, traveling structures, phase coexistence, and chaotic dynamics~\cite{you2020nonreciprocity, soto2014, dinelli2023non, duan2023dynamical}. At the continuum level, many of these phenomena can be captured by generalized Cahn--Hilliard-type theories incorporating nonreciprocal couplings between fields~\cite{saha2020scalar, frohoff2023nonreciprocal, you2020nonreciprocity}. Visually interacting active systems therefore provide a natural microscopic realization of nonreciprocal interactions, offering a simple yet powerful framework for understanding how asymmetric perception alone can drive collective motion not requiring velocity alignment or steric exclusion~\cite{strombom2011, barberis2016}.

The spatial extent over which agents perceive and respond to their neighbors is a fundamental ingredient of collective dynamics and naturally introduces a characteristic length scale into vision-based interactions. In this work, we investigate a visual perception-driven model of active particles interacting through purely position-based cohesive reorientation under a finite vision cone, following the framework introduced in ~\cite{barberis2016,peruani_2026}. By systematically varying the interaction range between the previously studied short-range and distance-independent limits, we explore how a finite perception horizon shapes collective self-organization. This question is particularly relevant because both natural and artificial agents typically sense and respond only to neighbors within a limited spatial extent. We show that the intermediate regime supports a rich spectrum of emergent states, including gas, nematic-band, and polar-file states, while also giving rise to previously unreported network-like structures sustained by dynamically evolving particle streams. Our results demonstrate that finite-range, non-reciprocal vision-based interactions provide a simple yet powerful mechanism for generating complex collective behavior in active systems.

\begin{figure*}[tbh!]
		\resizebox{0.8\textwidth}{!} {\includegraphics{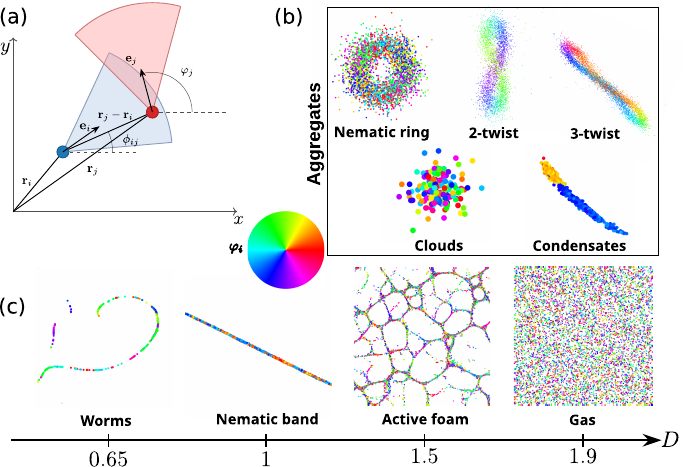}}
		\caption{\justifying (a) Perception-mediated interactions: The $j^\text{th}$ particle (red) lies within the vision cone of the $i^\text{th}$ particle (blue), causing the latter to reorient its propulsion direction towards the former. (b) Aggregates appear primarily at large vision angles and have correlation dimension $D = 0.1$--$0.3$. (c) Representative long-range phases, which appear primarily at smaller vision angles, characterized by the correlation dimension $D$: worms ($\theta = \pi/4$, $R_v = 6$), nematic bands ($\theta = \pi/6$, $R_v = 5$), active foams ($\theta = \pi/8$, $R_v = 5$), and a gas phase ($\theta = \pi/12$, $R_v = 1$). The other parameters are $\Phi = 1$, $v_0 = 0.1$ and $\Omega = 3$. The circular color wheel indicates the propulsion direction (orientation) of each particle.}
		\label{fig1}
\end{figure*}
 
\section{Methods}

The model consists of $N$ self-propelled point particles moving at a constant speed $v_0$ and interacting through a position-based vision-cone (VC) interaction that reorients their direction of self-propulsion towards their visible neighbors, thereby promoting cohesion and group formation. Since the interaction acts only within a restricted \emph{field of view}, given by a vision angle $\theta\le\pi$, of each particle, it is inherently anisotropic and introduces nonreciprocity in the picture. Here, all particles are assigned the same vision cone.

\subsection*{Equations of Motion}

In two dimensions, the state of particle $i$ is described by its position $\bm{r}_i$ and orientation angle $\varphi_i$. The dynamics are given by
\begin{equation}
    \label{eq1}
	m\ddot{\bm{r}}_i= -\gamma\dot{\bm{r}}_i + \gamma v_0 \,\bm{e}(\varphi_i) \,,
\end{equation}
where $\bm{e}(\varphi_i) = (\cos\varphi_i,\, \sin\varphi_i)^T$, and
\begin{equation}
\label{eq2}    
    \dot{\varphi}_i = \frac{\Omega}{n_i} \sum_{j \in VC_i} \sin(\alpha_{ij} - \varphi_i) + \sqrt{2D_r}\xi_i(t).
\end{equation}

Here $\bm{e}(\varphi_i)$ denotes the unit vector along the propulsion direction of the $i$-th particle, with $\varphi_i$ being its polar angle at time $t$, $v_0$ is the constant self-propulsion speed, and $\gamma$ is the translational friction coefficient. The angular dynamics comprises of a visual interaction-induced reorientation term and rotational noise. The noise $\xi_i(t)$ is Gaussian and white, with zero mean and unit variance, satisfying $\langle \xi_i(t)\rangle=0$ and $\langle \xi_i(t)\xi_j(t') \rangle = \delta_{ij}\delta(t-t')$.
The rotational diffusion coefficient $D_r$ sets the strength of this noise.

The interaction term is purely geometric: particle $i$ tends to align its orientation direction towards the relative position of the neighboring particles that it can \emph{see}. The angle $\alpha_{ij}$ is the polar angle of the unit vector $\bm{r}_{ij}/|\bm{r}_{ij}|$ which points from $i$th to $j$th particle, with $\bm{r}_{ij} = \bm{r}_j - \bm{r}_i$, as shown in Fig.~\ref{fig1}(a). The normalization factor $n_i$ (the number of neighbors within the visual zone of the $i$-th particle) ensures that the interaction strength remains non-additive. In absence of steric repulsion, multiple particles can occupy the same spatial location, leading to an arbitrarily large number of neighbors. This normalization thus ensures interaction strength remains finite and bounded, and prevents unphysical behaviors.

\subsection*{Vision Cone and Interaction Rule}

The set $VC_i$, at the right of Eq.~\ref{eq2}, contains all particles $j\text{ }(\ne i)$ that lie within the vision cone of the $i$th particle at any instant. The $j$th particle belongs to $VC_i$ if it satisfies both a distance and an angular constraint:
\begin{align}
	|\bm{r}_j - \bm{r}_i| &\leq R_v \,, \\
	\frac{\bm{r}_j - \bm{r}_i}{|\bm{r}_j - \bm{r}_i|} \cdot \bm{e}(\varphi_i) &> \cos(\theta) \, .
\end{align}

Here, $R_v$ defines the spatial extent of the vision cone or the vision range, while the vision angle $\theta$ is the half opening angle of the vision cone. For $\theta < \pi$, interactions are generically non-reciprocal, i.e., particle $j$ may influence $i$ without being influenced back.

\subsection*{Simulation Details}

Simulations are performed in a square domain of size $L\times L$ with periodic boundary conditions. The global density is fixed at $\Phi = N/L^2 = 1$, with $L = 100$ and the number of particles being $N = 10^4$ (unless stated otherwise). We have set $m=1$ and $\gamma=50$, giving the inertial time-scale $\tau_I=m/\gamma=2\times 10^{-2}$ and integrated equation~(\ref{eq1})  using the velocity-Verlet algorithm with a time step of $\Delta t = 5 \times 10^{-3}$. Unless stated otherwise, the rotational diffusion coefficient is fixed at $D_r=0.08$ (hence persistence time $\tau_p\simeq 1/D_r=12.5$, much larger than $\tau_I$), while the self-propulsion speed is set to $v_0=0.1$. In contrast to Refs.~\cite{barberis2016, peruani_2026}, where $D_r$ is varied as a control parameter, we fix $D_r$ at a small value and thereby work in the low-noise regime. This allows us to isolate the effect of cohesive interactions on the collective dynamics, free from confounding contributions from noise-induced transitions. By choosing appropriate characteristic length and time scales, the equations of motion can be nondimensionalized, allowing the corresponding scale-setting parameters to be fixed without loss of generality. Unless otherwise stated, all quantities are reported in simulation units.

The system is characterized by the following control parameters: the visual interaction strength $\Omega$ and the vision cone parameters $(R_v, \theta)$. Simulations are run for $1\times10^7$ time steps with the last $5\times 10^6$ timesteps used for averaging, discarding the initial transients.

\subsection*{Correlation (fractal) dimension} To quantitatively characterize the degree of spatial clustering and distinguish between different aggregation patterns, we employ the correlation dimension, $D$. This quantity is obtained from the spatial correlation function $C(r)$, which gives the probability that two particles are separated by a distance less than $r$:

\begin{equation}
    C(r) = \frac{2}{N(N-1)} \sum_{i=1}^{N} \sum_{j=i+1}^{N} \Theta(r - |\mathbf{r}_i - \mathbf{r}_j|)
\end{equation}
where $N$ is the total number of particles and $\Theta$ is the Heaviside step function. For a fractal structure, $C(r)$ follows the scaling $C(r) \sim r^{D}$. The correlation dimension is then extracted from the slope of the linear regime in a log-log plot~\cite{strogatz2001nonlinear},
\begin{equation}
    D = \frac{\log C(r)}{\log r}.
\end{equation}

A vanishing $D$ indicates highly localized point-like structures. As it increases towards unity, structures with one-dimensional morphology such as elongated, filamentary, lane, or band-like patterns, could be expected, while $D\sim2$ represents area-filling and homogeneous structures. The intermediate regime, $1\le D\le2$, corresponds to fractal structures, such as porous gels, labyrinthine patterns, and vascular-like networks, which extend across the plane without filling it uniformly~\cite{khuntia2025self, nasr2021supramolecular, lazzari2016fractal, co2019colloidal, cheeseman2022estimating}.

\section{Results}
The primary objective of this study is to explore the role of non-reciprocal interactions in determining structure formation in the minimal cognitive model introduced in Refs.~\cite{barberis2016,peruani2017hydrodynamic,peruani_2026}. We investigate this by systematically varying two key model parameters: the interaction range $(R_v)$ and the degree of
non-reciprocity $(\theta)$, and analyzing their impact on the resulting structural patterns.

\subsection{\label{phases}Emergent phases}

We classify the collective behaviors into four distinct categories: aggregates, worms, nematic bands, and active foams. Representative snapshots of these observed states are presented in Fig.~\ref{fig1}(b) and (c).

{\bf Aggregates:}
Aggregates comprise of a diverse collection of macroscopic clusters, including compact condensates, rings, diffusive clouds, and closed filamentary structures~\cite{peruani_2026}. Representative examples are shown in Fig.~\ref{fig1}(b). We observe closed filaments with nematic order (rings) that exhibit motion reminiscent of chiral active particles, while closed 8-shaped filaments with polar order containing a single defect (2-twist) maintain a net polarity and undergo ballistic motion. More complex closed filament structures containing two such defects, classified as 3-twist, are also observed. The net polarity of these structures vanishes, resulting in the absence of ballistic motion. However, they exhibit rotational motion due to the torque generated from the balance of forces. The last two structures typically coexist and dynamically interconvert, often through a transient open filament state showing strong translation (see movie-1, 2 \& 5 in the supplemental material (SM)~\cite{supplementary}). 

The classification and nomenclature adopted here follow Ref.~\cite{peruani_2026}, where defects are identified as singular topological points at which the filament crosses itself, creating discontinuities in the local polarization field. However, these defects are not topological in nature. Although they share structural similarities with $+2$ nematic topological defects, their cores remain nematically ordered, unlike conventional liquid-crystalline defects that possess isotropic cores~\cite{de1993physics}. Diffusive clouds exhibit no coherent collective motion due to the absence of net polarity. Condensates--highly localized structures containing $\sim\mathcal{O}(N)$ particles~\cite{wangCondensationSynchronizationAligning2024}--appear as a crossover between worm-like states and larger aggregates, and exhibit erratic motion (see movie-3 \& 4 in the SM~\cite{supplementary}). The transport properties of the aggregates are strongly coupled to their topology. Rings and other closed structures with twists has been observed in several vision-cone based studies~\cite{strombom2011, strombom2015shape, li2019role}. 

\begin{figure}[tbh!]
		\resizebox{\columnwidth}{!} {\includegraphics{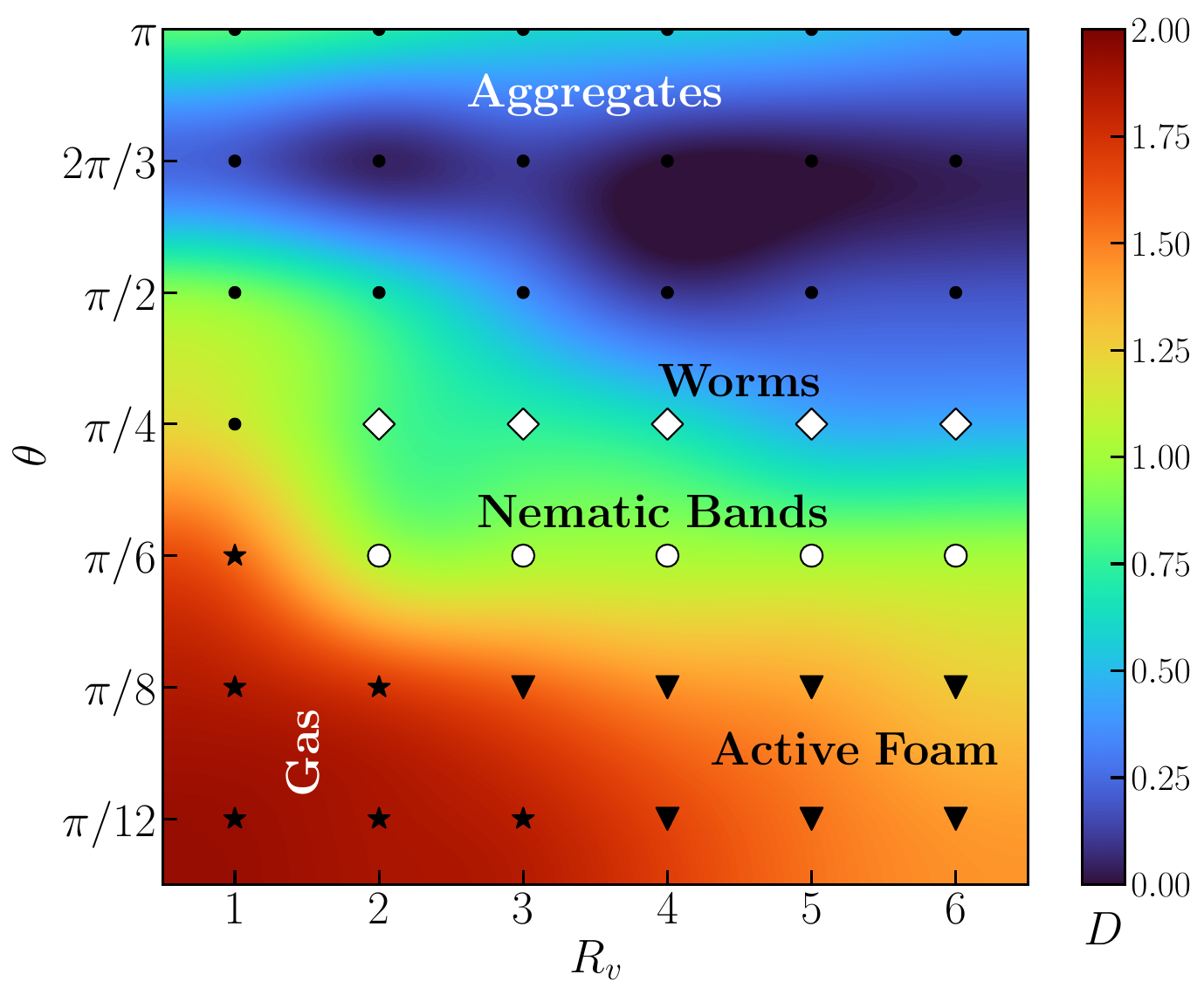}}
		\caption{\justifying Phase diagram of the observed collective states for $v_0 = 0.1$ and $\Omega = 3$. The vision range $R_v$ is varied along the $x$ axis, while the vision cone angle $\theta$ is varied along the $y$ axis. The heatmap shows the mean correlation fractal dimension $D$. Black circles: aggregates, stars: gas, white diamonds: worms, white circles: nematic bands, inverted triangle: active foams.}
		\label{fig2}
\end{figure}

{\bf Worms:}
Worms are cohesive, elongated, quasi-one-dimensional files, with weak polar order at the global scale [Fig.~\ref{fig1}(c)]. The non-interacting nature of the particles leads to pronounced back-and-forth motion of individual particles along the structure, giving rise to observable local nematicity, while the chain as a whole undergoes frequent fragmentation and merger (see movie-6 in the SM~\cite{supplementary}). Qualitatively similar worm-like states have been reported in point-particle systems~\cite{strombom2015shape, barberis2016, costanzo2018spontaneous, li2019role}, however, there is significant morphological difference with the worms found in systems with both vision-mediated and steric interactions~\cite{negi2022emergent, negi2024collective, bhaskar2026non}.

\begin{figure*}[tbh!]
    \centering
    \begin{subfigure}[t]{0.48\linewidth}
        \centering
        \includegraphics[width=\linewidth]{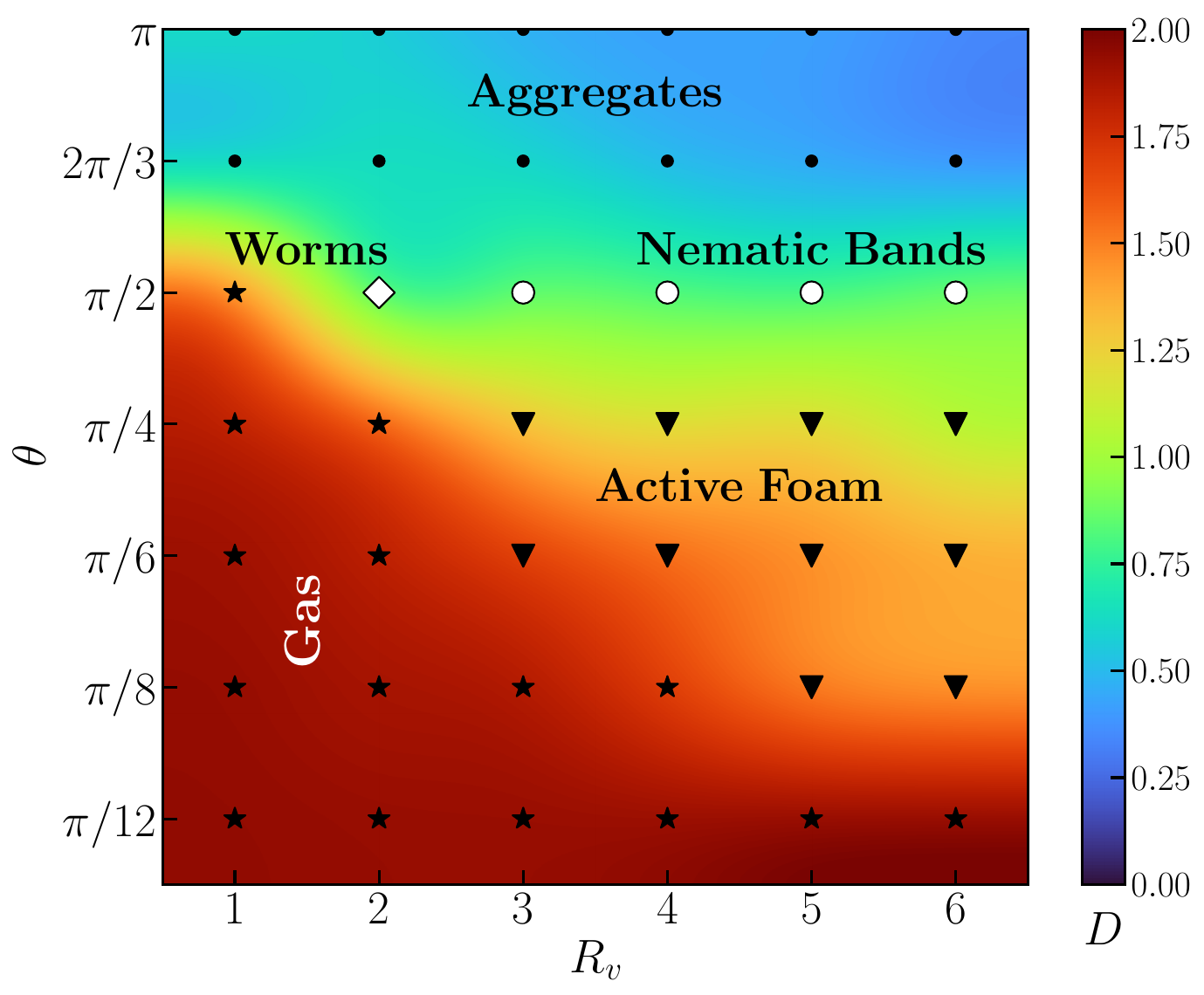}
        \caption{ $\Omega = 1$}
        \label{fig3a}
    \end{subfigure}
    \hfill
    \begin{subfigure}[t]{0.48\linewidth}
        \centering
        \includegraphics[width=\linewidth]{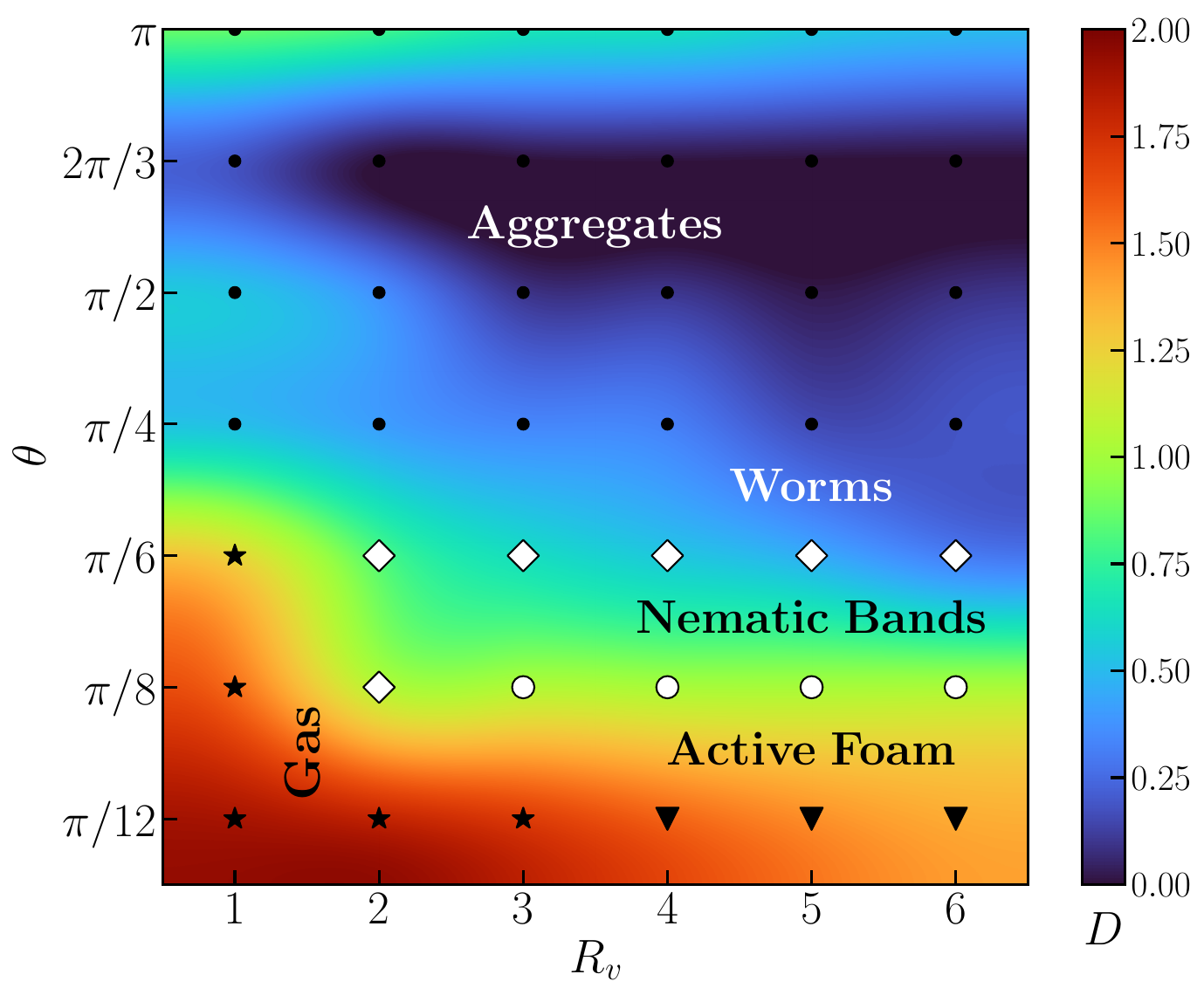}
        \caption{$\Omega = 5$}
        \label{fig3b}
    \end{subfigure}

    \caption{\justifying State diagrams of the observed collective states for $v_0=0.1$ at two different values of the visual torque strength $\Omega$. The vision range $R_v$ is varied along the $x$-axis, while the vision cone angle $\theta$ is varied along the $y$-axis. The heatmap represents the mean correlation fractal dimension. (a) $\Omega=1$. (b) $\Omega=5$. The heatmap shows the mean correlation fractal dimension $D$. Black circles: aggregates, stars: gas, white diamonds: worms, white circles: nematic bands, inverted triangle: active foams.}
    \label{fig:state_theta_Rv}
\end{figure*}

{\bf Nematic bands:}
Nematic bands are macroscopic, highly rectilinear structures exhibiting long-range nematic order~\cite{barberis2016,peruani_2026}. Following an initial transient during which smaller bands nucleate, grow, and merge, the system relaxes to a steady state consisting of one or several stable and near-static bands. Particle motion within a band is bidirectional, consistent with nematic symmetry (see movie-7 in the SM~\cite{supplementary}). It is observed that curved configurations at the transient state gradually relaxes to a perfectly straight one, suggesting the greater stability of the later structure. This could be due to specific combinations of $R_v$ and $\theta$ that generate a stabilizing transverse torque, thereby minimizing particle escape and maintaining the rectilinear morphology.

{\bf Active foams:}
Active foams are system-spanning, network-like structures with a fractal-like morphology and no global orientational order [Fig.~\ref{fig1}(c)]. The network consists of interconnected streams of particles, within which the particles are locally nematically aligned. These streams continuously branch, break, and reconnect, yet the large-scale network remains statistically stationary over long times. The state thus combines pronounced local rearrangements with persistent large-scale connectivity. This gives rise to spatially heterogeneous particle transport along streams and fluctuating foam-like domains, which show no systematic coarsening towards a homogeneous phase (see movie-8 in the SM~\cite{supplementary}).

\subsection{\label{pd1}State diagram}

Figure~\ref{fig2} summarizes the collective states observed in the $(\theta, R_v)$ parameter plane for a visual torque strength of $\Omega=3$. For lower values of the control parameters $\theta$ and $R_v$, the system remains in a disordered gaseous state, as each particle interacts with only a few neighbors and is therefore mostly isolated. As $\theta$ and $R_v$ increase slightly, these structures subsequently coarsen and merge, resulting in different macroscopic structures depending on the values of the control parameters. At small $\theta$, on increasing $R_v$ each particle acquires more neighbors, and the system initially forms chain-like (or line-like) structures with local nematic orientational order, consistent with the underlying cognitive interactions, and gradually leading to the transformation of these isolated chains into network-like structures that span the entire system, forming an active foam. On increasing $\theta$, active-foams become unstable, and the system transitions to a nematic band state due to slightly higher orientational influences. Further increase of $\theta$ destabilizes the nematic-bands which breaks up and reorganize into worm-like structures. For large $\theta$, approaching full reciprocity, the anisotropic states give way to aggregate phases, including condensates, rings, dynamic twisted structures, and diffusive clouds.

Overall, increasing $\theta$ drives a transition from highly directional, network-like structures to compact clustered states, while increasing $R_v$ generally promotes collective organization by enlarging the interaction neighborhood. The background heatmap in Fig.~\ref{fig2} shows the mean correlation dimension $D$, which provides a quantitative measure of morphology. Aggregates at very high VC shows $D\simeq 0.1-0.6$, as expected for highly concentrated or condensate-like states. Worms also show $D\leq1$, indicating a disconnected chain-like structure. Bands show $D\sim1$, pointing to its continuous, straight line-like configuration, as already discussed. The foam or network-like structure shows $D\simeq 1.5$, pointing to their partly-filled, more open structures. In colloidal studies, structures formed through diffusion-limited aggregation (DLA) or reaction-limited cluster-cluster aggregation (RLCA) are known to show similar $D$ values~\cite{co2019colloidal}. Finally, the homogeneous gas-like state approach the two-dimensional limit, $D\approx2$. Further justification for using $D$ as a quantitative marker is provided in the SM~\cite{supplementary}.

\subsubsection*{\label{pd2}Effect of maneuverability strength on the state diagram}

Since the visual torque strength, $\Omega$, introduces an interaction timescale that competes with the persistence time of individual particles, $1/D_r$, we examine how this competition affects the collective organization by constructing state diagrams for lower and higher torque strengths, namely $\Omega=1$ and $\Omega=5$. The resulting state diagrams are shown in Fig.~\ref{fig:state_theta_Rv}.

Two prominent trends can be identified from these state diagrams. First, increasing the maneuverability strength substantially enlarges the region occupied by aggregate states, which remain stable over a much wider range of $\theta$. For instance, at $\Omega=5$, aggregate states are observed even for narrow VCs down to $\theta=\pi/4$, whereas for smaller $\Omega$ they are largely confined to wider VCs. Thus, increasing $\Omega$ (hence reducing the characteristic timescale of visual reorientation), combined with moderate to large VC, prevents persistent (diffusive) random walk and promotes aggregation. Consequently, even a small number of neighbors within a particle's interaction zone can be sufficient to initiate and sustain group formation.

Second, the expansion of the aggregate region at higher $\Omega$ is accompanied by a suppression of the active-foam region, which becomes confined to a relatively narrow range of low $\theta$ values. At the same time, the nematic-band phase shifts markedly from larger $\theta$ at low $\Omega$ to smaller $\theta$ at high $\Omega$. This trend may be understood from the visual-torque term in Eq.~\ref{eq2}, $\sim \Omega\sum \sin(\alpha_{ij}-\varphi_i)$. As the vision cone narrows, only neighbors located closer to the particle's vision-cone axis contribute to the visual interaction, thereby reducing the average angular offset $(\alpha_{ij}-\varphi_i)$. For $(\alpha_{ij}-\varphi_i)<\pi/2$, the sine function increases monotonically, so a smaller angular offset results in a weaker visual torque for a fixed value of $\Omega$. Consequently, a larger torque strength is required to produce an equivalent reorientation effect, explaining why the nematic-band phase shifts toward lower $\theta$ as $\Omega$ increases. Furthermore, the stability of the worm phase becomes restricted to a narrow parameter regime, while cloud-like aggregates emerge upon increasing $\theta$ beyond the nematic-band region. At $\Omega=1$, however, the active-foam phase persists at relatively short $R_v$ for larger $\theta$, highlighting the subtle interplay among vision range, vision angle, and torque strength. We will address this in detail in the following.

\subsection{\label{dynamics}Dynamical measures}
{\bf Mean square displacement:}
To quantify the translational dynamics of the particles in the different collective states we calculate the mean square displacements (MSD), given as

\begin{equation}
\langle r^2(t)\rangle = \frac{1}{N}\sum_{i=1}^N \langle (\bm r_i(t+t_0) - \bm r_i(t_0))^2\rangle_{t_0}    .
\end{equation}

\begin{figure}[tbh!]
		\resizebox{\columnwidth}{!} {\includegraphics{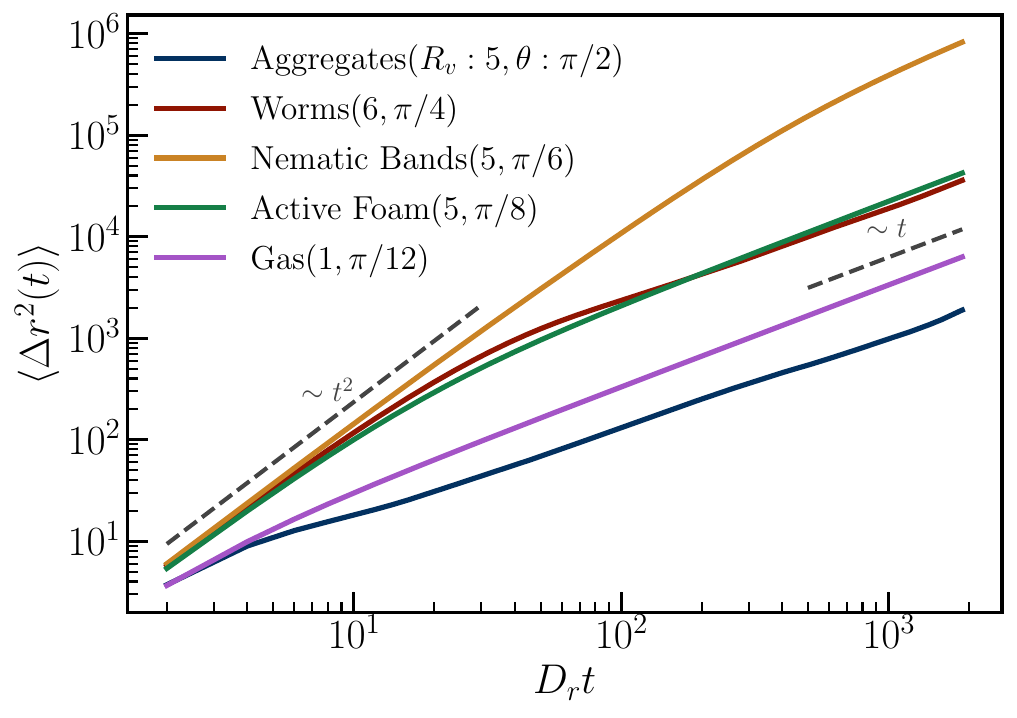}}
		\caption{\justifying Mean squared displacement at $v_0 = 0.1$ and $\Omega = 3$ for representative stable phases. Data are averaged over 5 independent realizations.}
		\label{fig4}
\end{figure}

Figure~\ref{fig4} shows the dimensionless MSD for the different collective phases at $v_0 = 0.1$ and $\Omega=3$ as a function of the nondimentionalized lag time $D_r t$. At early times $(2\lesssim D_rt\lesssim10)$, almost all phases exhibit ballistic behavior, $\langle \Delta r^2(t)\rangle \sim t^2$, because rotational diffusion has not yet randomized particle orientations and the particles retain their persistent self-propelled motion at speed $v_0$, as in non-interacting ABPs.

At intermediate- to late-times $(D_rt>10)$, clear state-dependent dynamics emerge. Particles in the nematic bands and worms display the largest MSD, remaining well above the gas phase over extended durations. The particles in the nematic-bands move predominantly along a one-dimensional path, exhibiting prolonged near-ballistic behavior; while those in the worms show an initial ballistic regime followed by a crossover to diffusion, which is a result of the local nematicity as described in Sec.~\ref{phases}. 

Particles in the active-foam phase exhibit intermediate transport behavior, with an MSD lying between those of the ordered phases (worms and bands) and the gaseous state. The crossover to diffusive dynamics occurs at shorter times than in worms or bands; as particles travel along interconnected streams, they repeatedly change direction at junctions and undergo local reorganization, which suppresses long-range persistent motion and limits transport at long times.

Particles in the gas-like state do not show noticeable clustering and tend to show normal active particle-like behavior, which is also reflected in the corresponding MSD. Initial super-diffusive scaling quickly crossovers to the diffusive one. Very short-ranged visual sensing, coupled with extremely narrow vision cone leaves the particles essentially blind.

In contrast, condensate-like aggregates show the smallest long-time MSD. Although initially ballistic, particles within dense condensates become confined, leading to subdiffusive behavior. Rings show chiral ABP-like motion; 2-twist structure shows ballistic motion because of the presence of single defect crossing whereas 3-twist structure shows rotational motion due to presence of double crossing which nullifies the net translational motion but generates a torque. We do not discuss them further in this study as some of those aspects are already discussed at other places~\cite{barberis2016,peruani_2026}.

All of the states show diffusive $(\langle \Delta r^2(t)\rangle \sim t)$ scaling at very long times, $D_rt\gg1$, which is expected, as a single active particle in this regime shows $\langle \Delta r^2(t)\rangle\sim 2v_0^2t/D_r$.\\

\begin{figure}[tbh!]
		\resizebox{\columnwidth}{!} {\includegraphics{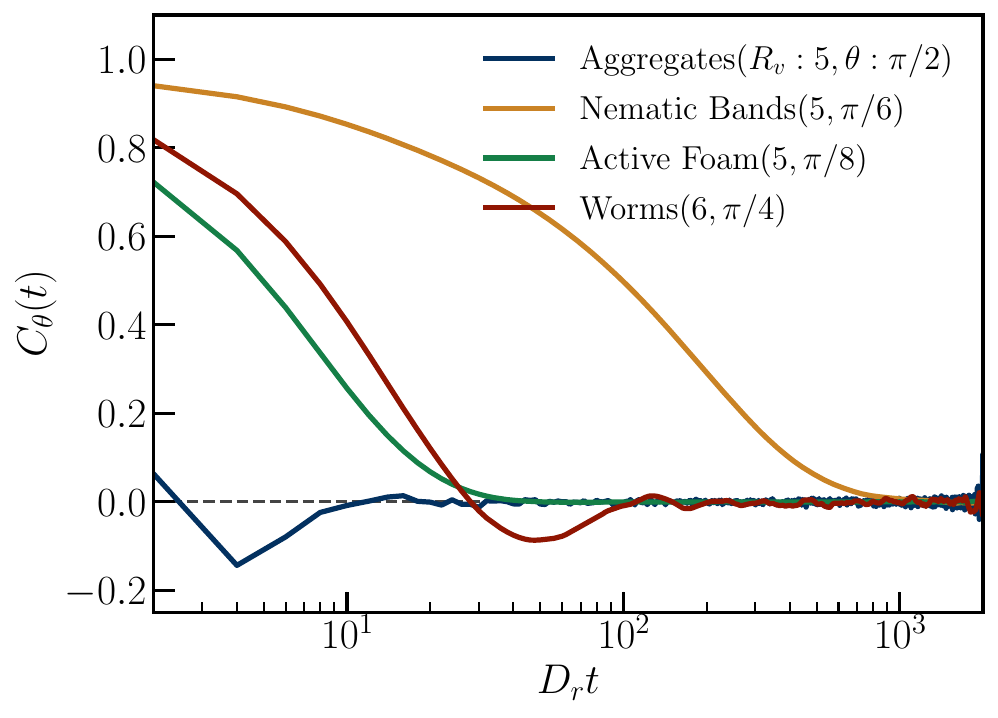}}
		\caption{\justifying Temporal Orientational Dynamics at $v_0 = 0.1$ and $\Omega = 3$ for representative stable phases. Data are averaged over 5 independent realizations. }
		\label{fig5}
\end{figure}

{\bf Temporal Orientational Dynamics:} To probe the orientational dynamics, we compute the temporal orientational correlation function, given as
\begin{equation}
    C_{\theta}(t) = \frac{1}{N} \sum_{i=1}^{N} \left\langle \bm{e}_i(t + t_0) \cdot \bm{e}_i(t_0) \right\rangle_{t_0},
\end{equation}
where $\bm{e}_i$ is the orientation of particle $i$ and $N$ is the total number of particles. $C_{\theta}$ is a measure of how persistent the orientational correlation of a particle is, as a function of the lag-time. The data for various phases is shown in Fig.~\ref{fig5}.

As discussed above, particles in the gaseous state exhibits almost no cognitive functionality. Consequently, there is no special orientational characteristic and $C_{\theta}(t)$ decays to zero very fast (not shown here).\\ 

For nematic bands, orientational memory is strongly preserved. The correlation remains high ($C_{\theta} > 0.8$) for several rotational time units, reflecting the stability of macroscopic bands in which neighbor interactions maintain alignment and suppress individual rotational fluctuations. Both the MSD and the OCF suggests that particles keep on moving in the same direction within the nematic-band. Particles in the mostly-transient worms keep on reverting their direction of motion. At the same time, the worms move about erratically, causing a gradual decay and the anti-correlation of $C_{\theta}$.

Particles in the active foam state shows rather fast decay of $C_{\theta}$. While in the curved branches, the particles move rather uninterrupted, showing the initial peak in the correlation. But on approaching the vertices, its subsequent orientation is uncertain. It may turn towards other branches or may even reverse its direction. This causes the quick decay.

Particles in the condensate-type aggregate state displays a markedly different behavior: $C_{\theta}(t)$ decays rapidly and subsequently shows anti-correlation. The particles are point-like and self-propel continuously, moving toward the highly concentrated condensate core but repeatedly overshooting it. Isotropic visual perception then causes them to turn back toward the core, although not necessarily along their previous trajectories. This back-and-forth motion together with the chaotic motion of the cluster as a whole, results onto the correlation pattern observed here. 

\subsection{\label{s1s2}Local Orientational Order}

\begin{figure*}[t]
    \centering
    \includegraphics[width=\linewidth]{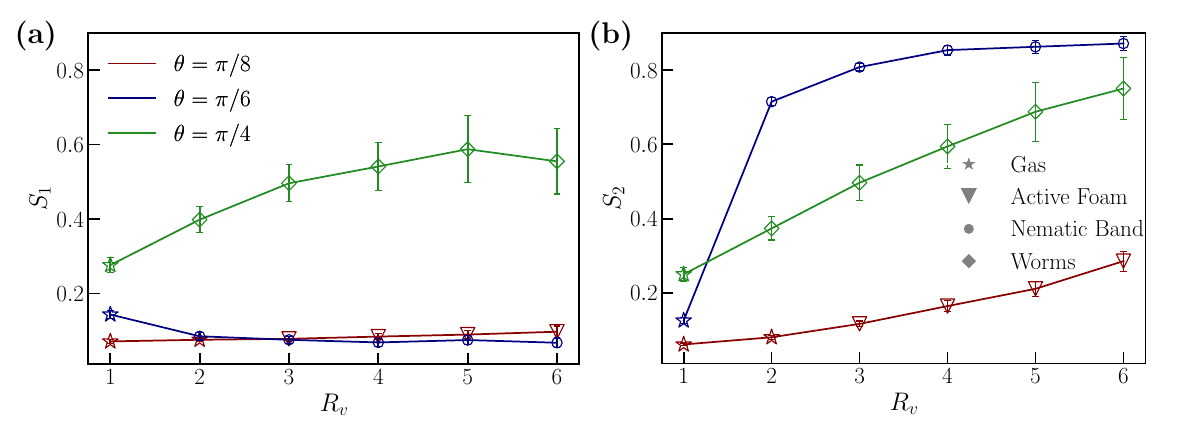}
    \caption{\justifying (a) Polar order parameter, $S_1$, as a function of the vision range $R_v$ for $\Omega=3$. (b) Nematic order parameter, $S_2$, as a function of the vision range $R_v$ for $\Omega=3$. The datasets correspond to vision angles $\theta=\pi/4$ (worms), $\theta=\pi/6$ (nematic bands), and $\theta=\pi/8$ (active foam). In all three cases, the lowest values of $R_v$ correspond to the gaseous state, indicated by star markers. The error bars represent the temporal fluctuations of the order parameters within each state. The comparatively larger error bars for the worm state reflect the fact that several worm-like aggregates continuously merge and split. Averaging is done over the last $1000$ frames.}
    \label{fig6}
\end{figure*}

To quantify the degree of orientational ordering in the system, we employ a \emph{local} orientational order parameter, defined as 

\begin{equation}
S_q=
\left|
\frac{1}{N}
\sum_{k=1}^{N}
e^{iq\theta_k}
\right|,
\label{eq9}
\end{equation}

where, for each particle, all other neighboring particles within a circular search radius of $r_c=10$ are identified. $N$ is the number of particles within the local neighborhood (including the reference particle), $\theta_k$ is the orientation angle of the $k$th neighbor, and $q$ is an integer symmetry index specifying the orientational symmetry being probed. The averaged order parameter, $\langle S_q\rangle$, is obtained by averaging $S_q$ over all particles and, where appropriate, over multiple steady-state configurations.

For $q=1$, Eq.~(\ref{eq9}) defines the local \textit{polar} order parameter, $S_1$, which measures the extent to which neighboring particles align and move in the same direction. $q=2$ defines the local \textit{nematic} order parameter, $S_2$, for which the factor of two in the exponential renders orientations separated by $\pi$ equivalent, thereby quantifying alignment along a common axis irrespective of the direction of motion. In both cases, values close to unity indicate strong orientational order, while values close to zero correspond to the absence of it.

Figures~\ref{fig6}(a,b) show the averaged polar and nematic order parameters as functions of the vision range $R_v$ for three vision angles. For $\theta=\pi/4$, the system transitions from a gaseous state at the smallest values of $R_v$ to the worm state as $R_v$ increases (Fig.~\ref{fig2}). For $\theta=\pi/6$, the gaseous state is likewise observed at the smallest $R_v$, followed by the nematic-band state. For $\theta=\pi/8$, the system exhibits an active-foam state over most of the explored range of $R_v$, while the smallest $R_v$ corresponds to the gaseous state. 

In the case of worms, for lower values of $R_v$, $S_1$ and $S_2$ are of comparable magnitude. As $R_v$ increases, $S_2$ becomes larger than $S_1$, indicating that the worms possess significant local nematic order while the polar order is comparatively weak. The nematic-bands display the highest degree of nematic ordering, followed by the worms. As in Fig.~(\ref{fig3a}), the transition from worm to nematic-bands on increasing $R_v$ points to a similar transition from a state with comparable polar--nematic ordering to one dominated by nematic order. It is worth noting that a state with a high polar order ($S_1$) necessarily possesses a high nematic order ($S_2$), but not vice versa. Beyond $R_v=4$, local nematic ordering in the worms start dominating. The active-foams also exhibits a finite nematic order, which becomes more pronounced with increasing vision range.

\subsection{Active foams}

\begin{figure*}
    \centering
    \includegraphics[width=\linewidth]{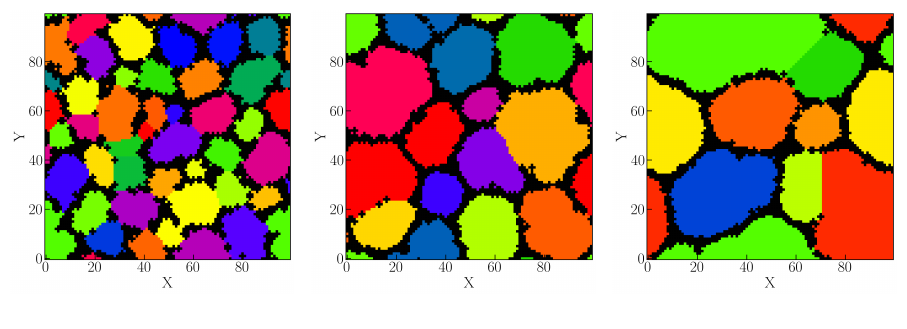}
    \caption{\justifying Voids in the active foam exhibiting a network-like morphology. From left to right, the parameters are $R_v = 5$, $7.5$, and $10$. Other parameters are fixed at $\theta=\pi/12$, $v_0 = 0.1$ and $\Omega = 3$. The colors indicate distinct voids only, while the black regions represent the particles. In some large voids, more than two colors appear. This is an artifact of the void-identification algorithm, which occasionally fails to assign black pixels to very narrow particle streams separating adjacent voids.}
    \label{fig7}
\end{figure*}

We dedicate this section to detailing the observed \emph{active foam} structure, characterized by a network-like morphology and fractal-like organization. To the best of our knowledge, such foam-like collective behavior has not been explicitly characterized in models of cognitive active agents. We therefore briefly review related instances of network-like structure formation reported in other active-matter systems. Foam-like morphologies have been observed in systems of nematically aligning point particles with orientational memory~\cite{chate_2015} and in systems with chirality disorder~\cite{ventejouSusceptibility2021}, although these studies did not specifically characterize or analyze the said state.

Related network-forming behavior has been reported in several active-matter systems. Examples include chiral active particles subjected to spatially correlated noise~\cite{fehlinger2025fluctuation}, microtubule--motor mixtures~\cite{sumino2012large,maryshev2020pattern}, and filamentous cyanobacteria exhibiting non-reciprocal alignment and trail-following interactions~\cite{faluweki2023active}. Dynamic networks have also been observed in C. elegans through humidity-mediated attraction and nematic alignment~\cite{sugi2019c}.

Network-like morphologies have also emerged in continuum models of active phase separation, including Active Model B+ and related theories~\cite{fausti2021capillary,gulati2025active}, as well as the Active Model H$^-$ framework, where a connected filamentous network accompanies arrested coarsening~\cite{lucas2026spontaneous}. Similar network formation is also well documented in biological processes such as vasculogenesis~\cite{merks2006cell,mehes2014collective}. These studies demonstrate that network morphologies can arise through diverse mechanisms, including active phase separation, non-reciprocal interactions, chemotaxis, and fluctuations.

Despite these similarities in morphology, the mechanisms underlying network formation in these systems differ from the visual-perception-driven mechanism considered here. In our model, active foams emerge solely from finite-range, anisotropic, non-reciprocal steering interactions, without explicit velocity alignment or physical volume exclusion. This provides a distinct route to the formation and maintenance of dynamic, system-spanning networks.


\subsubsection{\label{foammechanism}Mechanism of foam formation}

We focus here on the $\Omega=3$ case. As the vision angle $\theta$ decreases, the nematic-band state gives way to the active-foam state. This transition can be attributed to two associated effects of reducing $\theta$: (i) enhanced follower-like dynamics, arising from a narrower angular distribution of target particles, and (ii) an increased relative influence of rotational fluctuations. For a narrow vision cone, particles respond to a more restricted set of neighbors within their field of view, thereby weakening cohesive group formation. Consequently, even small fluctuations in particle orientations can progressively destabilize ordered structures by randomizing their subsequent directions of motion.

Starting either from a square-lattice configuration or from a spatially homogeneous random distribution, local accumulation driven by cohesive interactions, together with the follower dynamics, initially generates numerous loosely connected vertices and short edges. As the system evolves, these structures coarsen into more prominent yet dynamic vertices and edges. Rotational fluctuations appear to play a crucial role in selecting this dynamical regime. For sufficiently strong rotational noise, particles reorient too rapidly for persistent edges to develop. The system instead relaxes towards a microphase-separated state consisting of compact aggregate-like domains, which subsequently undergo slow coalescence over long time scales (see SM~\cite{supplementary} for the figures). In the opposite limit of very weak rotational noise, the persistence of particle motion becomes sufficiently strong so that only rapidly translating condensates form. At an intermediate level of rotational fluctuations, these two competing tendencies are balanced: particle trajectories are sufficiently persistent to sustain elongated edges, while fluctuations are also sufficient to promote local reorientation and the formation of vertices~\cite{noise}. The resulting dynamics resembles an arrested-coarsening scenario, in which the network does not fully coarsen into macroscopic domains but instead remains as a continuously restructuring, network-like active foam.

We further investigated this mechanism through a parameter-quenching. Starting from a fully developed nematic-band state, a sudden quench into the active-foam regime leads to the gradual transformation of the bands into a network-like structure. The reduced vision angle continues to support local nematic ordering; however, the transverse orientational torques (between particle-pairs maintaining the vision-cone criteria but with low $\alpha_{ij}$), become insufficient to maintain the rectilinear structures characteristic of the nematic bands. At the same time, the enhanced relative importance of orientational fluctuations facilitates branching away from the original bands. We observe such branching events to occur at a regular spatial interval along the bands, although a detailed analysis is beyond the scope of the present work. Repeated branching progressively generates vertices and edges and eventually fills the simulation domain. The resulting active foam remains dynamically reorganizing: edges repeatedly branch to generate new vertices and cells, while individual cells subsequently collapse through the contraction of their bounding edges (see Movie~8 in the SM~\cite{supplementary}). The reverse transformation is also observed. When a fully developed active foam is subjected to a sudden increase in the vision angle, the network gradually relaxes towards the nematic-band state. The formation of active foam on quenching from nematic bands, as well as on initializing in a square lattice or randomly, shows that it is not merely an artifact of the initial condition rather a robust distinct dynamical organizations.

\emph{Role of torque strength}: A decrease in the torque strength $\Omega$ weakens the cohesive interactions within the edges, leading to looser and wider edge structures. Consequently, a larger vision angle $\theta$ is required at lower $\Omega$ to sustain the local nematic-band-like ordering along the edges and prevent particles from spreading. At smaller $\theta$, enhanced orientational fluctuations promote further fragmentation of the edges, eventually leading to homogenization of the system. This trend is also evident from a comparison of the $\Omega=1$ and $\Omega=5$ phase diagrams in Fig.~\ref{fig:state_theta_Rv}, which shows that the active-foam regime extends to larger vision angles at $\Omega=1$. These observations highlight the role of the torque strength $\Omega$ in sustaining the cohesive network structure and, in turn, facilitating the formation of the active-foam state.

Limited finite-size analysis indicates the active-foams to persist across varying simulation domain, as well as upon reasonable density variation; see SM~\cite{supplementary} for details. 

\subsubsection{\label{voidarea}Structural characterization of active foams}
The structure of the active-foam state is characterized by a network of edges and vertices that encloses voids of varying sizes. Figure~\ref{fig7} shows the variation of the characteristic void (or cell) size with the vision range $R_v$ at fixed $\theta$ and $\Omega$. As the void size increases, the number of vertices decreases, with both quantities exhibiting a strong dependence on $R_v$. \\\\
\emph{Void area distribution}: 
Our observations indicate that, for a fixed VC, the sizes of the persistent yet dynamic voids enclosed by the edges and vertices of the active-foam network depend on the vision range $R_v$. Moreover, at any given time, the system exhibits a distribution of void areas rather than a single characteristic size. To quantify this behavior, we identify and characterize the void spaces using the SNOW (Sub-Network of an Over-segmented Watershed) algorithm as implemented in the PoreSpy library~\cite{PhysRevE.96.023307, gostick2019porespy}. The measurements are obtained by first averaging over the final 1,000 steady-state simulation frames and then performing an ensemble average over 10 independent realizations. Representative results are shown in Fig.~\ref{fig7} for three interaction ranges, $R_v=5 $, $7.5 $, and $10 $, at $v_0 = 0.1$ and vision maneuverability $\Omega=3$.

\begin{figure}
    \centering
    \includegraphics[width=\linewidth]{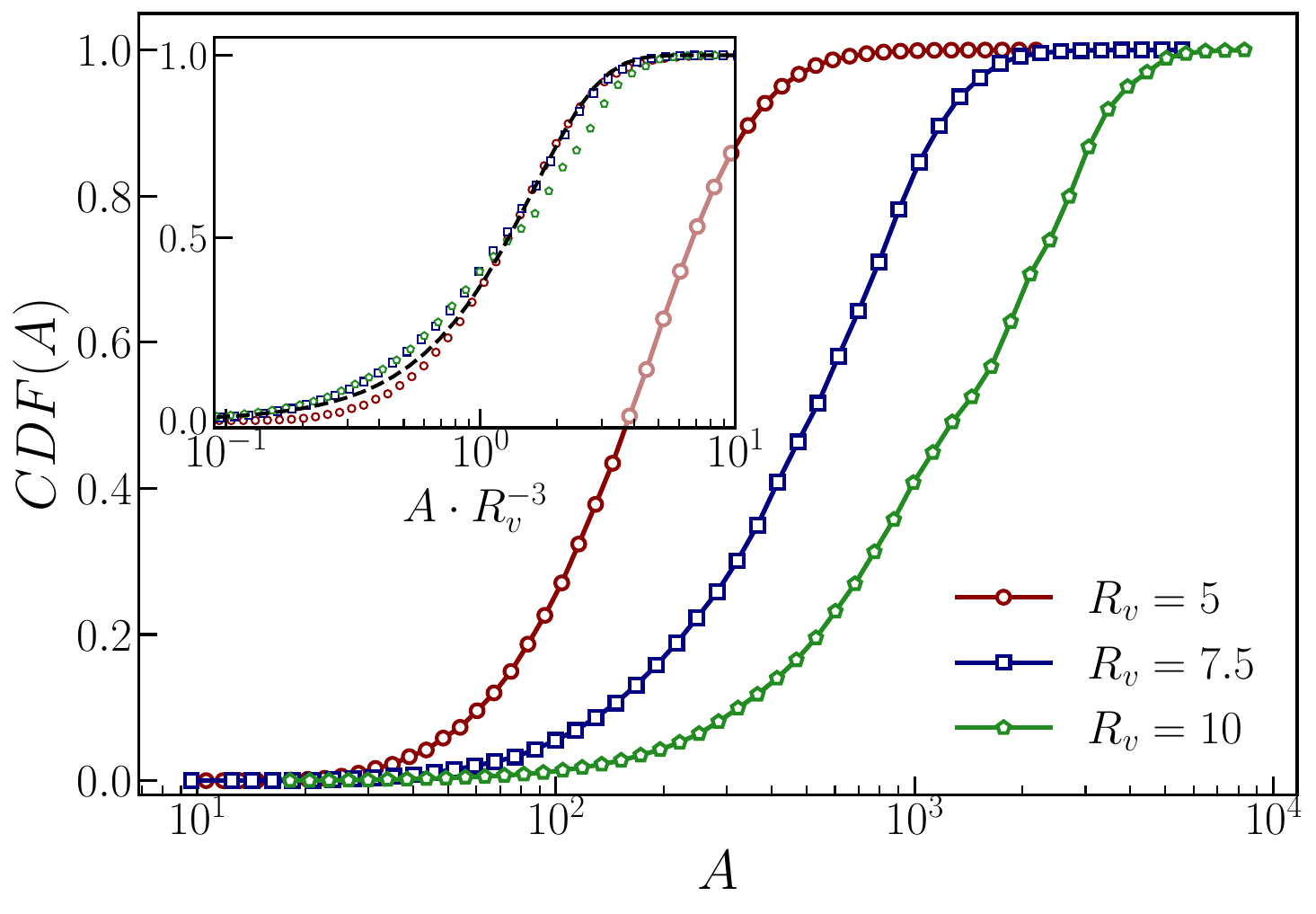}
    \caption{\justifying Cumulative distribution function of the void area in an active foam. Inset: Collapse of the CDF data as a function of the scaled quantity: $A/R_v^3$.}
    \label{fig8}
\end{figure}

The resulting cumulative distribution function of the void-area is shown in figure~\ref{fig8}. The curve for $R_v=5 $ exhibits a comparatively narrow spread in void area and saturates at smaller $A$, indicating that the voids possess relatively similar characteristic sizes with fewer large-area voids. Increasing $R_v$ leads to an increase in both the maximum void area and the dispersion of void sizes. For $R_v=10$, the minimum void size is also larger than in the other cases. The estimated maximum void areas are $\sim 2\times10^3$, $6\times10^3$, and $8\times10^3$, corresponding to equivalent radii of $\sim25$, $43$, and $50$ for $R_v=5, 7.5,\text{ and }10$, respectively. For still larger $R_v$, the system gradually evolves into a cross-shaped configuration consisting of two intersecting nematic bands (not shown).

As shown in Fig.~\ref{fig8}, the cumulative distribution function (CDF) of the void area $A$ rises more rapidly for smaller values of $R_v$, indicating a greater abundance of small voids. To examine the dependence on the vision range, we replot the CDF as a function of the rescaled variable $A/R_v^{\alpha}$. We find that $\alpha=3$ yields a satisfactory collapse of the CDFs for all investigated values of $R_v$ onto a single master curve (Fig.~\ref{fig8}, inset). This approximate collapse suggests that the dependence of the void-area distribution on the vision range can be largely absorbed into the single scaling variable $A/R_v^3$, while the functional form of the distribution remains nearly unchanged. Thus, increasing $R_v$ primarily rescales the overall structural features of the network without substantially modifying its statistical organization. At present, the exponent $\alpha=3$ should be regarded as an empirical scaling exponent inferred from the data collapse rather than from a theoretical prediction. Interestingly, the independently measured correlation dimension of the network is $D \approx 1.5$, such that $\alpha \approx 2D$. Whether this correspondence reflects a deeper connection between the fractal organization of the network and the scaling of the void statistics remains an open question.

\emph{Degree distribution}:
To characterize the topology of the active-foam network, we treat the foam vertices as network nodes and the edges connecting them as network edges. We analyze the distribution of the total number of vertices, $P(N_v)$, and the vertex-degree distribution, $P(k)$, which characterizes the local connectivity of the network~\cite{newman2007component,salas2025contact}. For smaller $R_v$, configurations with a larger number of vertices become more probable, consistent with the smaller characteristic void areas observed in Fig.~\ref{fig7}. As $R_v$ increases, the typical number of vertices decreases. The degree distribution further reveals that three-fold vertices ($k=3$) are the most probable, followed by four-fold vertices, while vertices with $k\geq5$ are essentially absent. Thus, despite changes in the characteristic void size with $R_v$, the network predominantly comprises three-fold and four-fold vertices, suggesting a geometrically constrained network topology~\cite{barthelemy2011spatial}. Further details of these distributions are provided in the Supplementary Material~\cite{supplementary}.

\subsubsection{\label{lengthscale}Emergent length scale of the network structure}

\begin{figure}
    \centering
    \includegraphics[width=\linewidth]{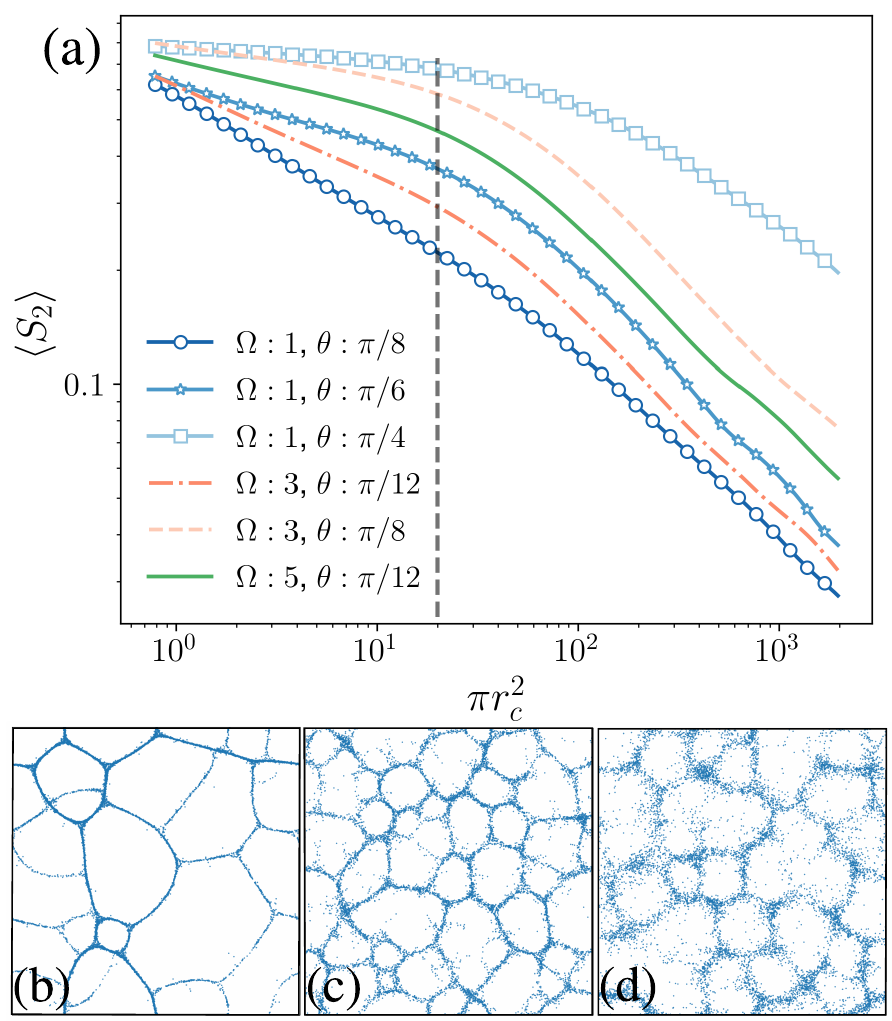}
    \caption{\justifying (a) Variation of the \emph{local} nematic order parameter $\langle S_2\rangle$ as a function of the search area $\pi r_c^2$. The vision range is fixed at $R_v=5$ in all cases. Averaging is done over all particles within the search area, and over the last $100$ frames. The dashed vertical line at $\pi r_c^2=20$ marks the cut-off search area $\pi r_c^{*2}$. (b)-(d) representative snapshots for $\Omega=1$ and $\theta=\pi/4$, $\pi/6$, and $\pi/8$, respectively.}
    \label{fig9}
\end{figure}

To investigate whether a characteristic length scale in the active-foam systems can be found, we again evaluate Eq.~\ref{eq9} to find the local nematic ordering $\langle S_2\rangle$ in the system, while systematically varying the search radius $r_c$--or the search area $\pi r_c^2$ from small to large values. The result is shown in Fig.~\ref{fig9}(a).

The uppermost curve, corresponding to $\Omega=1$ and $\theta=\pi/4$, exhibits two distinct regimes: a plateau in $\langle S_2\rangle$ up to a crossover search radius $r_c=r_c^*$, followed by a decay at larger $r_c$. The plateau indicates that local nematic order persists within the network structure, most prominently along the edges, whereas the subsequent decay reflects the loss of such local order. For $\Omega=1$ and $\theta=\pi/6$, the plateau becomes less pronounced and is replaced by a slow decay, followed by a crossover to a more rapid decrease beyond $\sim r_c^*$. At $\theta=\pi/8$, $\langle S_2\rangle$ exhibits an approximately power-law-like decay, indicating the absence of a well-defined characteristic length scale as in a disordered state~\cite{nishiguchi2017long, faluweki2023active}. This structural disorder is also evident in the snapshots of the active foams shown in Fig.~\ref{fig9}(b)-(d).

The curves corresponding to other values of $\Omega$ exhibit qualitatively similar behavior. However, the spatial range over which significant local nematic order persists progressively decreases with decreasing $\theta$. For the active-foam structure at $\Omega=1$ and $\theta=\pi/4$, this order persists, on average, up to a search area of $\pi r_c^2\approx20$, corresponding to $r_c\approx2.5$; slightly larger than the persistence length $l_p=v_0/D_r=1.25$. 

The observed trend is remarkably similar to that observed in Ref.~\cite{faluweki2023active}, where the diminishing plateau corresponded to a decreasing density of the filamentous cyanobacteria.

The complementary question regarding the effect of vision range $R_v$ on the emergent length scale, keeping other parameters fixed, is briefly discussed in the SM~\cite{supplementary}.

\section{Conclusions and Discussions}

Biologically inspired vision-mediated interactions provide a fundamentally different mechanism for collective motion than conventional metric or topological alignment. In this work, we investigated a system of visually interacting active particles by extending the cognitive flocking model of Refs.~\cite{barberis2016, peruani_2026} to incorporate a finite vision range, $R_v$, and systematically explored the combined effects of $R_v$ and the vision angle $\theta$. This extension is motivated by the largely unexplored role of sensing range in vision-based active matter, despite the fact that biological and synthetic agents exhibit widely varying perception ranges. Our study reveals a rich state diagram comprising gaseous, worm, nematic-band, aggregate, and, most notably, an \emph{active-foam} state. 

Increasing either the vision angle or the vision range enlarges the effective interaction neighborhood, driving the system from gaseous behavior toward increasingly ordered collective states. At small values of $\theta$ and $R_v$, particles remain in the gaseous state, whereas intermediate values of $\theta$ give rise to worm and nematic-band phases characterized by distinct translational and order parameters. Sufficiently large vision angles stabilize aggregates and narrow ones lead to extended active-foam networks at finite $R_v$. These phases are quantitatively distinguished by their correlation dimension $D$.

The different collective phases also exhibit distinct transport and relaxation dynamics. While the gaseous state displays the characteristic dynamics of isolated active particles, nematic-bands exhibit long-lived ballistic motion arising from persistent orientational coherence. Aggregate states are predominantly subdiffusive, reflecting the confinement of particles around the cluster center of mass, whereas particles in the active-foam state display intermediate transport with an earlier crossover to normal diffusion. Temporal correlation functions further reveal slow orientational relaxation in the ordered phases and distinct signatures associated with the rotational dynamics of aggregates. Finally, increasing the visual torque strength $\Omega$ primarily enhances aggregation, while its decrease stabilizes gaseous and foam states. 

Our quench experiments demonstrate reversible transitions between the active-foam and nematic-band states upon crossing their respective stability boundaries, indicating that these phases occupy neighboring regions of parameter space. In foams, the gradual increase in the characteristic void size with $R_v$ further indicates structures consisting of intersecting streams of oppositely moving particles, while the structure as a whole undergoes continuous reorganization. We identify an emergent length scale that depends strongly on both the vision range and vision angle. Our network-based analysis further suggests that the active-foam morphology is characterized by a geometrically constrained network topology. The persistence of this morphology across variations in system size and particle density demonstrates its robustness, while its structural organization is strongly modulated by the strength of rotational noise.

Similar network-like states have also been reported in experimental and theoretical studies on \emph{C. elegans}~\cite{sugi2019c} or cyanobacteria~\cite{faluweki2023active}, microtubule--motor mixtures~\citep{sumino2012large,chate_2015}. Experiments revealed lattices composed of hexagonally arranged vortices connected by nematically aligned microtubules, whereas theoretical studies incorporating orientational memory predicted a dynamically stable active-foam phase. Likewise, analytical studies of compressible active mixtures of microtubules and molecular motors found nematic vortex lattice and foam-like patterns in the limit of weak elasticity and strong tangential anchoring, where density gradients strongly influence orientational order~\cite{maryshev2020pattern}. Related foam-like states have also been observed in dry nematically aligning active systems with chirality disorder~\cite{ventejouSusceptibility2021}. Lastly, theoretical models related to active phase separation are increasingly discovering network-like morphological formation~\cite{fausti2021capillary, gulati2025active, lucas2026spontaneous}. Active foam is even being considered as a new type of phase separation along with bulk phase separation, microphase and bubbly phase separation~\cite{fausti2021capillary}.

Active foams therefore appear to be a robust collective state arising from effective interaction anisotropy rather than any specific interaction mechanism. Identifying the minimal ingredients required for their emergence remains an important open question. Future work could explore more realistic distance-dependent vision interactions, three-dimensional systems, and the effects of steric interactions, confinement, and boundaries, while further assessing the relevance of active foams to biological collectives and autonomous agent design.\\

\nocite{10.1039/c5sm02350e}

\section*{Acknowledgments}

The authors thank Sudeshna Sinha (Mohali) for valuable discussions on fractal analysis and Dipak Patra (Mohali) for carefully reading the manuscript. The authors also acknowledge the computing facilities provided by IISER Mohali.\\

\section*{Author Contributions}
A.K.D. conceived and designed the research. A.D. performed the simulations. A.D. and D.B. analyzed the data. All authors contributed to the interpretation of the results. D.B. and A.K.D. wrote the manuscript.\\

\section*{Conflicts of interest}
There are no conflicts to declare.\\

\section*{Data availability}
The data associated with the figures is available at:
\url{https://github.com/Aryan-360/Minimal_Cognitive_Flocking_Model}\\


%


\begin{thebibliography}{73}
\bibitem [{noi()}]{noise}%
  \BibitemOpen
  \href@noop {} {}\bibinfo {note} {Note that the magnitude of the rotational
  noise directly influences the self-propulsion or P\'e through the relation
  Pe$\sim v_0/D_rL$, where $L$ is some length scale in the system such as the
  system size.}\BibitemShut {Stop}%
\end{thebibliography}
\end{document}